

%
%

%
%
%
%

\def\Serif{cmr}
\def\SerifBold{cmbx}
\def\SerifItalics{cmti}
\def\SerifSlanted{cmsl}
\def\SerifBoldItalics{cmbxti}
\def\SansSerif{cmss}
\def\SansSerifBold{cmssbx}
\def\SansSerifItalics{cmssi}
\def\SansSerifSlanted{cmssi}
\def\Math{cmmi}
\def\Symbols{cmsy}
\def\MathBold{cmmib}
\def\MoreSymbols{cmex}
\def\Typewriter{cmtt}
\def\Gothic{eufm}
\def\Double{msbm}
\def\Relazioni{msam}

= 			\Serif10 			at 5pt
= 		\SerifBold10 		at 5pt
= 	\SerifItalics10 	at 5pt
=		\SerifSlanted10 	at 5pt
=	\SerifBoldItalics10	at 5pt
= 		\SansSerif10 		at 5pt
=	\SansSerifBold10	at 5pt
=	\SansSerifItalics10	at 5pt
=	\SansSerifSlanted10	at 5pt
=				\Math10				at 5pt
=			\MathBold10			at 5pt
=			\Symbols10			at 5pt
=		\MoreSymbols10		at 5pt
=		\Typewriter10		at 5pt
=			\Gothic10			at 5pt
=			\Double10			at 5pt

= 			\Serif10 			at 7pt
= 		\SerifBold10 		at 7pt
= 	\SerifItalics10 	at 7pt
=	\SerifSlanted10 	at 7pt
=\SerifBoldItalics10	at 7pt
= 		\SansSerif10 		at 7pt
= 	\SansSerifBold10 	at 7pt
=\SansSerifItalics10	at 7pt
=\SansSerifSlanted10	at 7pt
=			\Math10				at 7pt
=		\MathBold10			at 7pt
=			\Symbols10			at 7pt
=		\MoreSymbols10		at 7pt
=		\Typewriter10		at 7pt
=			\Gothic10			at 7pt
=			\Double10			at 7pt

= 			\Serif10 			at 8pt
= 		\SerifBold10 		at 8pt
= 	\SerifItalics10 	at 8pt
=	\SerifSlanted10 	at 8pt
=\SerifBoldItalics10	at 8pt
= 		\SansSerif10 		at 8pt
= 	\SansSerifBold10 	at 8pt
=\SansSerifItalics10 at 8pt
=\SansSerifSlanted10 at 8pt
=			\Math10				at 8pt
=		\MathBold10			at 8pt
=			\Symbols10			at 8pt
=		\MoreSymbols10		at 8pt
=		\Typewriter10		at 8pt
=			\Gothic10			at 8pt
=			\Double10			at 8pt

= 			\Serif10 			at 10pt
= 		\SerifBold10 		at 10pt
= 		\SerifItalics10 	at 10pt
=		\SerifSlanted10 	at 10pt
=	\SerifBoldItalics10	at 10pt
= 		\SansSerif10 		at 10pt
= 	\SansSerifBold10 	at 10pt
= 	\SansSerifItalics10 at 10pt
= 	\SansSerifSlanted10 at 10pt
=				\Math10				at 10pt
=			\MathBold10			at 10pt
=			\Symbols10			at 10pt
=		\MoreSymbols10		at 10pt
=		\Typewriter10		at 10pt
=			\Gothic10			at 10pt
=			\Double10			at 10pt
=			\Relazioni10			at 10pt

= 				\Serif10 			at 12pt
= 			\SerifBold10 		at 12pt
= 		\SerifItalics10 	at 12pt
=		\SerifSlanted10 	at 12pt
=	\SerifBoldItalics10	at 12pt
= 			\SansSerif10 		at 12pt
= 		\SansSerifBold10 	at 12pt
= 	\SansSerifItalics10 at 12pt
= 	\SansSerifSlanted10 at 12pt
=				\Math10				at 12pt
=			\MathBold10			at 12pt
=			\Symbols10			at 12pt
=		\MoreSymbols10		at 12pt
=			\Typewriter10		at 12pt
=				\Gothic10			at 12pt
=				\Double10			at 12pt

= 			\Serif10 			at 14pt
= 		\SerifBold10 		at 14pt
= 	\SerifItalics10 	at 14pt
=		\SerifSlanted10 	at 14pt
=	\SerifBoldItalics10	at 14pt
= 		\SansSerif10 		at 14pt
= 	\SansSerifBold10 	at 14pt
= \SansSerifSlanted10 at 14pt
= \SansSerifItalics10 at 14pt
=				\Math10				at 14pt
=			\MathBold10			at 14pt
=			\Symbols10			at 14pt
=		\MoreSymbols10		at 14pt
=		\Typewriter10		at 14pt
=			\Gothic10			at 14pt
=			\Double10			at 14pt

\def\NormalStyle{\parindent=5pt\parskip=3pt\normalbaselineskip=14pt%
\def\nt{\tenSerif}%
\def\rm{\fam0\tenSerif}%
\textfont0=\tenSerif\scriptfont0=\sevenSerif\scriptscriptfont0=\fiveSerif
\textfont1=\tenMath\scriptfont1=\sevenMath\scriptscriptfont1=\fiveMath
\textfont2=\tenSymbols\scriptfont2=\sevenSymbols\scriptscriptfont2=\fiveSymbols
\textfont3=\tenMoreSymbols\scriptfont3=\sevenMoreSymbols\scriptscriptfont3=\fiveMoreSymbols
\textfont\itfam=\tenSerifItalics\def\it{\fam\itfam\tenSerifItalics}%
\textfont\slfam=\tenSerifSlanted\def\sl{\fam\slfam\tenSerifSlanted}%
\textfont\ttfam=\tenTypewriter\def\tt{\fam\ttfam\tenTypewriter}%
\textfont\bffam=\tenSerifBold%
\def\bf{\fam\bffam\tenSerifBold}\scriptfont\bffam=\sevenSerifBold\scriptscriptfont\bffam=\fiveSerifBold%
\def\cal{\tenSymbols}%
\def\greekbold{\tenMathBold}%
\def\gothic{\tenGothic}%
\def\Bbb{\tenDouble}%
\def\LieFont{\tenSerifItalics}%
\nt\normalbaselines\baselineskip=14pt%
}

\def\TitleStyle{\parindent=0pt\parskip=0pt\normalbaselineskip=15pt%
\def\nt{\fourteenSansSerifBold}%
\def\rm{\fam0\fourteenSansSerifBold}%
\textfont0=\fourteenSansSerifBold\scriptfont0=\tenSansSerifBold\scriptscriptfont0=\eightSansSerifBold
\textfont1=\fourteenMath\scriptfont1=\tenMath\scriptscriptfont1=\eightMath
\textfont2=\fourteenSymbols\scriptfont2=\tenSymbols\scriptscriptfont2=\eightSymbols
\textfont3=\fourteenMoreSymbols\scriptfont3=\tenMoreSymbols\scriptscriptfont3=\eightMoreSymbols
\textfont\itfam=\fourteenSansSerifItalics\def\it{\fam\itfam\fourteenSansSerifItalics}%
\textfont\slfam=\fourteenSansSerifSlanted\def\sl{\fam\slfam\fourteenSerifSansSlanted}%
\textfont\ttfam=\fourteenTypewriter\def\tt{\fam\ttfam\fourteenTypewriter}%
\textfont\bffam=\fourteenSansSerif%
\def\bf{\fam\bffam\fourteenSansSerif}\scriptfont\bffam=\tenSansSerif\scriptscriptfont\bffam=\eightSansSerif%
\def\cal{\fourteenSymbols}%
\def\greekbold{\fourteenMathBold}%
\def\gothic{\fourteenGothic}%
\def\Bbb{\fourteenDouble}%
\def\LieFont{\fourteenSerifItalics}%
\nt\normalbaselines\baselineskip=15pt%
}

\def\PartStyle{\parindent=0pt\parskip=0pt\normalbaselineskip=15pt%
\def\nt{\fourteenSansSerifBold}%
\def\rm{\fam0\fourteenSansSerifBold}%
\textfont0=\fourteenSansSerifBold\scriptfont0=\tenSansSerifBold\scriptscriptfont0=\eightSansSerifBold
\textfont1=\fourteenMath\scriptfont1=\tenMath\scriptscriptfont1=\eightMath
\textfont2=\fourteenSymbols\scriptfont2=\tenSymbols\scriptscriptfont2=\eightSymbols
\textfont3=\fourteenMoreSymbols\scriptfont3=\tenMoreSymbols\scriptscriptfont3=\eightMoreSymbols
\textfont\itfam=\fourteenSansSerifItalics\def\it{\fam\itfam\fourteenSansSerifItalics}%
\textfont\slfam=\fourteenSansSerifSlanted\def\sl{\fam\slfam\fourteenSerifSansSlanted}%
\textfont\ttfam=\fourteenTypewriter\def\tt{\fam\ttfam\fourteenTypewriter}%
\textfont\bffam=\fourteenSansSerif%
\def\bf{\fam\bffam\fourteenSansSerif}\scriptfont\bffam=\tenSansSerif\scriptscriptfont\bffam=\eightSansSerif%
\def\cal{\fourteenSymbols}%
\def\greekbold{\fourteenMathBold}%
\def\gothic{\fourteenGothic}%
\def\Bbb{\fourteenDouble}%
\def\LieFont{\fourteenSerifItalics}%
\nt\normalbaselines\baselineskip=15pt%
}

\def\ChapterStyle{\parindent=0pt\parskip=0pt\normalbaselineskip=15pt%
\def\nt{\fourteenSansSerifBold}%
\def\rm{\fam0\fourteenSansSerifBold}%
\textfont0=\fourteenSansSerifBold\scriptfont0=\tenSansSerifBold\scriptscriptfont0=\eightSansSerifBold
\textfont1=\fourteenMath\scriptfont1=\tenMath\scriptscriptfont1=\eightMath
\textfont2=\fourteenSymbols\scriptfont2=\tenSymbols\scriptscriptfont2=\eightSymbols
\textfont3=\fourteenMoreSymbols\scriptfont3=\tenMoreSymbols\scriptscriptfont3=\eightMoreSymbols
\textfont\itfam=\fourteenSansSerifItalics\def\it{\fam\itfam\fourteenSansSerifItalics}%
\textfont\slfam=\fourteenSansSerifSlanted\def\sl{\fam\slfam\fourteenSerifSansSlanted}%
\textfont\ttfam=\fourteenTypewriter\def\tt{\fam\ttfam\fourteenTypewriter}%
\textfont\bffam=\fourteenSansSerif%
\def\bf{\fam\bffam\fourteenSansSerif}\scriptfont\bffam=\tenSansSerif\scriptscriptfont\bffam=\eightSansSerif%
\def\cal{\fourteenSymbols}%
\def\greekbold{\fourteenMathBold}%
\def\gothic{\fourteenGothic}%
\def\Bbb{\fourteenDouble}%
\def\LieFont{\fourteenSerifItalics}%
\nt\normalbaselines\baselineskip=15pt%
}

\def\SectionStyle{\parindent=0pt\parskip=0pt\normalbaselineskip=13pt%
\def\nt{\twelveSansSerifBold}%
\def\rm{\fam0\twelveSansSerifBold}%
\textfont0=\twelveSansSerifBold\scriptfont0=\eightSansSerifBold\scriptscriptfont0=\eightSansSerifBold
\textfont1=\twelveMath\scriptfont1=\eightMath\scriptscriptfont1=\eightMath
\textfont2=\twelveSymbols\scriptfont2=\eightSymbols\scriptscriptfont2=\eightSymbols
\textfont3=\twelveMoreSymbols\scriptfont3=\eightMoreSymbols\scriptscriptfont3=\eightMoreSymbols
\textfont\itfam=\twelveSansSerifItalics\def\it{\fam\itfam\twelveSansSerifItalics}%
\textfont\slfam=\twelveSansSerifSlanted\def\sl{\fam\slfam\twelveSerifSansSlanted}%
\textfont\ttfam=\twelveTypewriter\def\tt{\fam\ttfam\twelveTypewriter}%
\textfont\bffam=\twelveSansSerif%
\def\bf{\fam\bffam\twelveSansSerif}\scriptfont\bffam=\eightSansSerif\scriptscriptfont\bffam=\eightSansSerif%
\def\cal{\twelveSymbols}%
\def\bg{\twelveMathBold}%
\def\gothic{\twelveGothic}%
\def\Bbb{\twelveDouble}%
\def\LieFont{\twelveSerifItalics}%
\nt\normalbaselines\baselineskip=13pt%
}

\def\SubSectionStyle{\parindent=0pt\parskip=0pt\normalbaselineskip=13pt%
\def\nt{\twelveSansSerifItalics}%
\def\rm{\fam0\twelveSansSerifItalics}%
\textfont0=\twelveSansSerifItalics\scriptfont0=\eightSansSerifItalics\scriptscriptfont0=\eightSansSerifItalics%
\textfont1=\twelveMath\scriptfont1=\eightMath\scriptscriptfont1=\eightMath%
\textfont2=\twelveSymbols\scriptfont2=\eightSymbols\scriptscriptfont2=\eightSymbols%
\textfont3=\twelveMoreSymbols\scriptfont3=\eightMoreSymbols\scriptscriptfont3=\eightMoreSymbols%
\textfont\itfam=\twelveSansSerif\def\it{\fam\itfam\twelveSansSerif}%
\textfont\slfam=\twelveSansSerifSlanted\def\sl{\fam\slfam\twelveSerifSansSlanted}%
\textfont\ttfam=\twelveTypewriter\def\tt{\fam\ttfam\twelveTypewriter}%
\textfont\bffam=\twelveSansSerifBold%
\def\bf{\fam\bffam\twelveSansSerifBold}\scriptfont\bffam=\eightSansSerifBold\scriptscriptfont\bffam=\eightSansSerifBold%
\def\cal{\twelveSymbols}%
\def\greekbold{\twelveMathBold}%
\def\gothic{\twelveGothic}%
\def\Bbb{\twelveDouble}%
\def\LieFont{\twelveSerifItalics}%
\nt\normalbaselines\baselineskip=13pt%
}

\def\AuthorStyle{\parindent=0pt\parskip=0pt\normalbaselineskip=14pt%
\def\nt{\tenSerif}%
\def\rm{\fam0\tenSerif}%
\textfont0=\tenSerif\scriptfont0=\sevenSerif\scriptscriptfont0=\fiveSerif
\textfont1=\tenMath\scriptfont1=\sevenMath\scriptscriptfont1=\fiveMath
\textfont2=\tenSymbols\scriptfont2=\sevenSymbols\scriptscriptfont2=\fiveSymbols
\textfont3=\tenMoreSymbols\scriptfont3=\sevenMoreSymbols\scriptscriptfont3=\fiveMoreSymbols
\textfont\itfam=\tenSerifItalics\def\it{\fam\itfam\tenSerifItalics}%
\textfont\slfam=\tenSerifSlanted\def\sl{\fam\slfam\tenSerifSlanted}%
\textfont\ttfam=\tenTypewriter\def\tt{\fam\ttfam\tenTypewriter}%
\textfont\bffam=\tenSerifBold%
\def\bf{\fam\bffam\tenSerifBold}\scriptfont\bffam=\sevenSerifBold\scriptscriptfont\bffam=\fiveSerifBold%
\def\cal{\tenSymbols}%
\def\greekbold{\tenMathBold}%
\def\gothic{\tenGothic}%
\def\Bbb{\tenDouble}%
\def\LieFont{\tenSerifItalics}%
\nt\normalbaselines\baselineskip=14pt%
}


\def\AbstractStyle{\parindent=0pt\parskip=0pt\normalbaselineskip=12pt%
\def\nt{\eightSerif}%
\def\rm{\fam0\eightSerif}%
\textfont0=\eightSerif\scriptfont0=\sevenSerif\scriptscriptfont0=\fiveSerif
\textfont1=\eightMath\scriptfont1=\sevenMath\scriptscriptfont1=\fiveMath
\textfont2=\eightSymbols\scriptfont2=\sevenSymbols\scriptscriptfont2=\fiveSymbols
\textfont3=\eightMoreSymbols\scriptfont3=\sevenMoreSymbols\scriptscriptfont3=\fiveMoreSymbols
\textfont\itfam=\eightSerifItalics\def\it{\fam\itfam\eightSerifItalics}%
\textfont\slfam=\eightSerifSlanted\def\sl{\fam\slfam\eightSerifSlanted}%
\textfont\ttfam=\eightTypewriter\def\tt{\fam\ttfam\eightTypewriter}%
\textfont\bffam=\eightSerifBold%
\def\bf{\fam\bffam\eightSerifBold}\scriptfont\bffam=\sevenSerifBold\scriptscriptfont\bffam=\fiveSerifBold%
\def\cal{\eightSymbols}%
\def\greekbold{\eightMathBold}%
\def\gothic{\eightGothic}%
\def\Bbb{\eightDouble}%
\def\LieFont{\eightSerifItalics}%
\nt\normalbaselines\baselineskip=12pt%
}

\def\RefsStyle{\parindent=0pt\parskip=0pt%
\def\nt{\eightSerif}%
\def\rm{\fam0\eightSerif}%
\textfont0=\eightSerif\scriptfont0=\sevenSerif\scriptscriptfont0=\fiveSerif
\textfont1=\eightMath\scriptfont1=\sevenMath\scriptscriptfont1=\fiveMath
\textfont2=\eightSymbols\scriptfont2=\sevenSymbols\scriptscriptfont2=\fiveSymbols
\textfont3=\eightMoreSymbols\scriptfont3=\sevenMoreSymbols\scriptscriptfont3=\fiveMoreSymbols
\textfont\itfam=\eightSerifItalics\def\it{\fam\itfam\eightSerifItalics}%
\textfont\slfam=\eightSerifSlanted\def\sl{\fam\slfam\eightSerifSlanted}%
\textfont\ttfam=\eightTypewriter\def\tt{\fam\ttfam\eightTypewriter}%
\textfont\bffam=\eightSerifBold%
\def\bf{\fam\bffam\eightSerifBold}\scriptfont\bffam=\sevenSerifBold\scriptscriptfont\bffam=\fiveSerifBold%
\def\cal{\eightSymbols}%
\def\greekbold{\eightMathBold}%
\def\gothic{\eightGothic}%
\def\Bbb{\eightDouble}%
\def\LieFont{\eightSerifItalics}%
\nt\normalbaselines\baselineskip=10pt%
}



%
%


\def\ModeYes{yes}
\def\ModeNo{no}

\def\ModeUndef{undefined}


\def\nx{\noexpand}
\def\ni{\noindent}
\def\newpage{\vfill\eject}

\def\ss{\vskip 5pt}
\def\ms{\vskip 10pt}
\def\bs{\vskip 20pt}

 \def\,{\mskip\thinmuskip}
 \def\!{\mskip-\thinmuskip}
 \def\>{\mskip\medmuskip}
 \def\;{\mskip\thickmuskip}

%
% ... 
%

\def\refsModePost{post}
\def\refsModeAuto{auto}

\def\dbRefsSatusModeOk{ok}
\def\dbRefsSatusModeError{error}
\def\dbRefsSatusModeWarning{warning}


\newcount\BNUM
\BNUM=0

\def\refs{}

\def\SetModePost{\xdef\refsMode{\refsModePost}}			
\SetModePost

\def\dbRefsStatusOk{%
	\xdef\dbRefsStatus{\dbRefsSatusModeOk}%
	\xdef\dbRefsError{\ModeNo}%
	\xdef\dbRefsWarning{\ModeNo}%
	\xdef\dbRefsInfo{\ModeNo}%
}

\def\dbRefs{%
}

\def\dbRefsGet#1{%
	\xdef\found{N}\xdef\ikey{#1}\dbRefsStatusOk%
	\xdef\key{\ModeUndef}\xdef\tag{\ModeUndef}\xdef\tail{\ModeUndef}%
	\dbRefs%
}

\def\NextRefsTag{%
	\global\advance\BNUM by 1%
}
\def\ShowTag#1{{\bf [#1]}}

\def\dbRefsInsert#1#2{%
\dbRefsGet{#1}%
\if\found Y %
   \xdef\dbRefsStatus{\dbRefsSatusModeWarning}%
   \xdef\dbRefsWarning{record is already there}%
   \xdef\dbRefsInfo{record not inserted}%
\else%
   \toks2=\expandafter{\dbRefs}%
   \ifx\refsMode\refsModeAuto \NextRefsTag
    \xdef\dbRefs{%
   	\the\toks2 \nx\xdef\nx\dbx{#1}%
	\nx\ifx\nx\ikey %
		\nx\dbx\nx\xdef\nx\found{Y}%
		\nx\xdef\nx\key{#1}%
		\nx\xdef\nx\tag{\the\BNUM}%
		\nx\xdef\nx\tail{#2}%
	\nx\fi}%
	\global\xdef\refs{\refs \ss\ni[\the\BNUM]\ #2\par}
   \fi%
   \ifx\refsMode\refsModePost 
    \xdef\dbRefs{%
   	\the\toks2 \nx\xdef\nx\dbx{#1}%
	\nx\ifx\nx\ikey %
		\nx\dbx\nx\xdef\nx\found{Y}%
		\nx\xdef\nx\key{#1}%
		\nx\xdef\nx\tag{\ModeUndef}%
		\nx\xdef\nx\tail{#2}%
	\nx\fi}%
   \fi%
\fi%
}

\def\dbRefsEdit#1#2#3{\dbRefsGet{#1}%
\if\found N 
   \xdef\dbRefsStatus{\dbRefsSatusModeError}%
   \xdef\dbRefsError{record is not there}%
   \xdef\dbRefsInfo{record not edited}%
\else%
   \toks2=\expandafter{\dbRefs}%
   \xdef\dbRefs{\the\toks2%
   \nx\xdef\nx\dbx{#1}%
   \nx\ifx\nx\ikey\nx\dbx %
	\nx\xdef\nx\found{Y}%
	\nx\xdef\nx\key{#1}%
	\nx\xdef\nx\tag{#2}%
	\nx\xdef\nx\tail{#3}%
   \nx\fi}%
\fi%
}

\def\bib#1#2{\RefsStyle\dbRefsInsert{#1}{#2}%
	\ifx\dbRefsStatus\dbRefsSatusModeWarning %
		\message{^^J}%
		\message{WARNING: Reference [#1] is doubled.^^J}%
	\fi%
}

\def\ref#1{\dbRefsGet{#1}%
\ifx\found N %
  \message{^^J}%
  \message{ERROR: Reference [#1] unknown.^^J}%
  \ShowTag{??}%
\else%
	\ifx\tag\ModeUndef \NextRefsTag%
		\dbRefsEdit{#1}{\the\BNUM}{\tail}%
		\dbRefsGet{#1}%
		\global\xdef\refs{\refs \ss\ni [\tag]\ \tail\par}
	\fi
	\ShowTag{\tag}%
\fi%
}

\def\ShowBiblio{\ms\Ensure{\SectionEnsure}%
{\SectionStyle\ni References}%
{\RefsStyle\refs}%
}

\newcount\CHANGES
\CHANGES=0
\def\AuxFile{7}
\def\PreventDoubleOn{\xdef\PreventDoubleLabel{\ModeYes}}

\PreventDoubleOn

\def\StoreLabel#1#2{\xdef\itag{#2}
 \ifx\PreModeStatus\ModeNo %
   \message{^^J}%
   \errmessage{You can't use Check without starting with OpenPreMode (and finishing with ClosePreMode)^^J}%
 \else%
   \immediate\write\AuxFile{\nx\dbLabelPreInsert{#1}{\itag}}%
   \dbLabelGet{#1}%
   \ifx\itag\tag %
   \else%
	\global\advance\CHANGES by 1%
 	\xdef\itag{(?.??)}%
    \fi%
   \fi%
}

\def\PreModeStatus{\ModeNo}

\def\ClosePreMode{\immediate\closeout\AuxFile%
  \ifnum\CHANGES=0%
	\message{^^J}%
	\message{**********************************^^J}%
	\message{**  NO CHANGES TO THE AuxFile  **^^J}%
	\message{**********************************^^J}%
 \else%
	\message{^^J}%
	\message{**************************************************^^J}%
	\message{**  PLAEASE TYPESET IT AGAIN (\the\CHANGES)  **^^J}%
    \errmessage{**************************************************^^ J}%
  \fi%
  \edef\PreModeStatus{\ModeNo}%
}

\def\dbLabelSatusModeOk{ok}

\def\dbLabelSatusModeWarning{warning}

\def\dbLabelStatusOk{%
	\xdef\dbLabelStatus{\dbLabelSatusModeOk}%
	\xdef\dbLabelError{\ModeNo}%
	\xdef\dbLabelWarning{\ModeNo}%
	\xdef\dbLabelInfo{\ModeNo}%
}

\def\dbLabel{%
}

\def\dbLabelGet#1{%
	\xdef\found{N}\xdef\ikey{#1}\dbLabelStatusOk%
	\xdef\key{\ModeUndef}\xdef\tag{\ModeUndef}\xdef\pre{\ModeUndef}%
	\dbLabel%
}

\def\ShowLabel#1{%
 \dbLabelGet{#1}%
 \ifx\tag \ModeUndef %
 	\global\advance\CHANGES by 1%
 	(?.??)%
 \else%
 	\tag%
 \fi%
}

\def\dbLabelPreInsert#1#2{\dbLabelGet{#1}%
\if\found Y %
  \xdef\dbLabelStatus{\dbLabelSatusModeWarning}%
   \xdef\dbLabelWarning{Label is already there}%
   \xdef\dbLabelInfo{Label not inserted}%
   \message{^^J}%
   \errmessage{Double pre definition of label [#1]^^J}%
\else%
   \toks2=\expandafter{\dbLabel}%
    \xdef\dbLabel{%
   	\the\toks2 \nx\xdef\nx\dbx{#1}%
	\nx\ifx\nx\ikey %
		\nx\dbx\nx\xdef\nx\found{Y}%
		\nx\xdef\nx\key{#1}%
		\nx\xdef\nx\tag{#2}%
		\nx\xdef\nx\pre{\ModeYes}%
	\nx\fi}%
\fi%
}

\def\dbLabelInsert#1#2{\dbLabelGet{#1}%
\xdef\itag{#2}%
\dbLabelGet{#1}%
\if\found Y %
	\ifx\tag\itag %
	\else%
	   \ifx\PreventDoubleLabel\ModeYes %
		\message{^^J}%
		\errmessage{Double definition of label [#1]^^J}%
	   \else%
		\message{^^J}%
		\message{Double definition of label [#1]^^J}%
	   \fi%
	\fi%
   \xdef\dbLabelStatus{\dbLabelSatusModeWarning}%
   \xdef\dbLabelWarning{Label is already there}%
   \xdef\dbLabelInfo{Label not inserted}%
\else%
   \toks2=\expandafter{\dbLabel}%
    \xdef\dbLabel{%
   	\the\toks2 \nx\xdef\nx\dbx{#1}%
	\nx\ifx\nx\ikey %
		\nx\dbx\nx\xdef\nx\found{Y}%
		\nx\xdef\nx\key{#1}%
		\nx\xdef\nx\tag{#2}%
		\nx\xdef\nx\pre{\ModeNo}%
	\nx\fi}%
\fi%
}


\newcount\PART
\newcount\CHAPTER
\newcount\SECTION
\newcount\SUBSECTION
\newcount\FNUMBER

\PART=0
\CHAPTER=0
\SECTION=0
\SUBSECTION=0	
\FNUMBER=0

\def\LastPart{\ModeUndef}
\def\LastChapter{\ModeUndef}
\def\LastSection{\ModeUndef}
\def\LastSubSection{\ModeUndef}
\def\LastClaim{\ModeUndef}
\def\Last{\ModeUndef}

\newdimen\TOBOTTOM
\newdimen\LIMIT

\def\Ensure#1{\ \par\ \immediate\LIMIT=#1\immediate\TOBOTTOM=\the\pagegoal\advance\TOBOTTOM by -\pagetotal%
\ifdim\TOBOTTOM<\LIMIT\newpage \else%
\vskip-\parskip\vskip-\parskip\vskip-\baselineskip\fi}

\def\PartLabel{\the\PART}
\def\NewPart#1{\global\advance\PART by 1%
         \bs\ni{\PartStyle  Part \PartLabel:}
         \bs\ni{\PartStyle #1}\newpage%
         \CHAPTER=0\SECTION=0\SUBSECTION=0\FNUMBER=0%
         \gdef\Left{#1}%
         \global\edef\Last{\PartLabel}%
         \global\edef\LastPart{\PartLabel}%
         \global\edef\LastChapter{\ModeUndef}%
         \global\edef\LastSection{\ModeUndef}%
         \global\edef\LastSubSection{\ModeUndef}%
         \global\edef\LastClaim{\ModeUndef}}
\def\ChapterLabel{\the\CHAPTER}
\def\NewChapter#1{\global\advance\CHAPTER by 1%
         \bs\ni{\ChapterStyle  Chapter \ChapterLabel: #1}\ms%
         \SECTION=0\SUBSECTION=0\FNUMBER=0%
         \gdef\Left{#1}%
         \global\edef\Last{\ChapterLabel}%
         \global\edef\LastChapter{\ChapterLabel}%
         \global\edef\LastSection{\ModeUndef}%
         \global\edef\LastSubSection{\ModeUndef}%
         \global\edef\LastClaim{\ModeUndef}}
\def\SectionEnsure{3cm}
\def\NewSection#1{\Ensure{\SectionEnsure}\gdef\SectionLabel{\the\SECTION}\global\advance\SECTION by 1%
         \ms\ni{\SectionStyle  \SectionLabel.\ #1}\ss%
         \SUBSECTION=0\FNUMBER=0%
         \gdef\Left{#1}%
         \global\edef\Last{\SectionLabel}%
         \global\edef\LastSection{\SectionLabel}%
         \global\edef\LastSubSection{\ModeUndef}%
         \global\edef\LastClaim{\ModeUndef}}
\def\NewAppendix#1#2{\Ensure{\SectionEnsure}\gdef\SectionLabel{#1}\global\advance\SECTION by 1%
         \bs\ni{\SectionStyle  Appendix \SectionLabel.\ #2}\ss%
         \SUBSECTION=0\FNUMBER=0%
         \gdef\Left{#2}%
         \global\edef\Last{\SectionLabel}%
         \global\edef\LastSection{\SectionLabel}%
         \global\edef\LastSubSection{\ModeUndef}%
         \global\edef\LastClaim{\ModeUndef}}
\def\Acknowledgements{\Ensure{\SectionEnsure}\gdef\SectionLabel{}%
         \ms\ni{\SectionStyle  Acknowledgments}\ss%
         \SECTION=0\SUBSECTION=0\FNUMBER=0%
         \gdef\Left{}%
         \global\edef\Last{\ModeUndef}%
         \global\edef\LastSection{\ModeUndef}%
         \global\edef\LastSubSection{\ModeUndef}%
         \global\edef\LastClaim{\ModeUndef}}
\def\SubSectionEnsure{2cm}
\def\SubSectionLabel{\ifnum\SECTION>0 \the\SECTION.\fi\the\SUBSECTION}
\def\NewSubSection#1{\Ensure{\SubSectionEnsure}\global\advance\SUBSECTION by 1%
         \ms\ni{\SubSectionStyle #1}\ss%
         \global\edef\Last{\SubSectionLabel}%
         \global\edef\LastSubSection{\SubSectionLabel}}
\def\SetNumberingModeN{\def\ClaimLabel{(\the\FNUMBER)}}
\def\SetNumberingModeSN{\def\ClaimLabel{(\ifnum\SECTION>0 \SectionLabel.\fi%
      \the\FNUMBER)}}
\def\SetNumberingModeCSN{\def\ClaimLabel{(\ifnum\CHAPTER>0 \the\CHAPTER.\fi%
      \ifnum\SECTION>0 \SectionLabel.\fi%
      \the\FNUMBER)}}

\def\NewClaim{\global\advance\FNUMBER by 1%
    \ClaimLabel%
    \global\edef\LastClaim{\ClaimLabel}%
    \global\edef\Last{\ClaimLabel}}

\def\HideLabels{\xdef\ShowLabelsMode{\ModeNo}}
\HideLabels

\def\fn{\eqno{\NewClaim}} 
\def\fl#1{%
\ifx\ShowLabelsMode\ModeYes%
 \eqno{{\buildrel{\hbox{\AbstractStyle[#1]}}\over{\hfill\NewClaim}}}%
\else%
 \eqno{\NewClaim}%
\fi%
\dbLabelInsert{#1}{\ClaimLabel}}
\def\fprel#1{\global\advance\FNUMBER by 1\StoreLabel{#1}{\ClaimLabel}%
\ifx\ShowLabelsMode\ModeYes%
\eqno{{\buildrel{\hbox{\AbstractStyle[#1]}}\over{\hfill.\itag}}}%
\else%
 \eqno{\itag}%
\fi%
}

\def\cl#1{\global\advance\FNUMBER by 1\dbLabelInsert{#1}{\ClaimLabel}%
\ifx\ShowLabelsMode\ModeYes%
${\buildrel{\hbox{\AbstractStyle[#1]}}\over{\hfill\ClaimLabel}}$%
\else%
  $\ClaimLabel$%
\fi%
}
\def\cprel#1{\global\advance\FNUMBER by 1\StoreLabel{#1}{\ClaimLabel}%
\ifx\ShowLabelsMode\ModeYes%
${\buildrel{\hbox{\AbstractStyle[#1]}}\over{\hfill.\itag}}$%
\else%
  $\itag$%
\fi%
}

\def\Note{\ms\leftskip 3cm\rightskip 1.5cm\AbstractStyle}
\def\endNote{\par\leftskip 2cm\rightskip 0cm\NormalStyle\ss}


\parindent=7pt
\leftskip=2cm
\newcount\SideIndent
\newcount\SideIndentTemp
\SideIndent=0
\newdimen\SectionIndent
\SectionIndent=-8pt

\def\sidebar{\vrule height15pt width.2pt }
\def\endcorner{\hbox{\hbox{\vrule height6pt width.2pt}\vbox to6pt{\vfill\hbox
to4pt{\leaders\hrule height0.2pt\hfill}}}}
\def\begincorner{\hbox{\hbox{\vrule height6pt width.2pt}\vbox to6pt{\hbox
to4pt{\leaders\hrule height0.2pt\hfill}}}}
\def\endbegincorner{\hbox{\vbox to15pt{\endcorner\vskip-6pt\begincorner\vfill}}}
\def\SideShow{\SideIndentTemp=\SideIndent \ifnum \SideIndentTemp>0 
\loop\sidebar\hskip 2pt \advance\SideIndentTemp by-1\ifnum \SideIndentTemp>1 \repeat\fi}

\def\BeginSection{{\vbadness 100000 \par\ni\hskip\SectionIndent%
\SideShow\vbox to 15pt{\vfill\begincorner}}\global\advance\SideIndent by1\vskip-10pt}

\def\EndSection{{\vbadness 100000 \par\ni\global\advance\SideIndent by-1%
\hskip\SectionIndent\SideShow\vbox to15pt{\endcorner\vfill}\vskip-10pt}}

\def\EndBeginSection{{\vbadness 100000\par\ni%
\global\advance\SideIndent by-1\hskip\SectionIndent\SideShow
\vbox to15pt{\vfill\endbegincorner}}%
\global\advance\SideIndent by1\vskip-10pt}

\def\ShowBeginCorners#1{%
\SideIndentTemp =#1 \advance\SideIndentTemp by-1%
\ifnum \SideIndentTemp>0 %
\vskip-15truept\hbox{\kern 2truept\vbox{\hbox{\begincorner}%
\ShowBeginCorners{\SideIndentTemp}\vskip-3truept}}%
\fi%
}

\def\ShowEndCorners#1{%
\SideIndentTemp =#1 \advance\SideIndentTemp by-1%
\ifnum \SideIndentTemp>0 %
\vskip-15truept\hbox{\kern 2truept\vbox{\hbox{\endcorner}%
\ShowEndCorners{\SideIndentTemp}\vskip 2truept}}%
\fi%
}

\def\BeginSections#1{{\vbadness 100000 \par\ni\hskip\SectionIndent%
\SideShow\vbox to 15pt{\vfill\ShowBeginCorners{#1}}}\global\advance\SideIndent by#1\vskip-10pt}

\def\EndSections#1{{\vbadness 100000 \par\ni\global\advance\SideIndent by-#1%
\hskip\SectionIndent\SideShow\vbox to15pt{\vskip15pt\ShowEndCorners{#1}\vfill}\vskip-10pt}}

\def\EndBeginSections#1#2{{\vbadness 100000\par\ni%
\global\advance\SideIndent by-#1%
\hbox{\hskip\SectionIndent\SideShow\kern-2pt%
\vbox to15pt{\vskip15pt\ShowEndCorners{#1}\vskip4pt\ShowBeginCorners{#2}}}}%
\global\advance\SideIndent by#2\vskip-10pt}




%
%


\def\al{\alpha}
\def\be{\beta}
\def\de{\delta}

\def\ep{\epsilon}

\def\te{\theta}
\def\la{\lambda}

\def\om{\omega}

\def\ka{\kappa}

\def\Ga{\Gamma}






\def\det{{\hbox{det}}}

\def\ip{\hbox to4pt{\leaders\hrule height0.3pt\hfill}\vbox to8pt{\leaders\vrule width0.3pt\vfill}\kern 2pt}
 
\def\del{\partial}
\def\na{\nabla}

\def\then{\Rightarrow}

%
%

\def\cases#1{\left\{\eqalign{#1}\right.}
\NormalStyle
\SetNumberingModeSN
\PreventDoubleOn

\long\def\title#1{\centerline{\TitleStyle\ni#1}}

\long\def\author#1{\ms\centerline{\AuthorStyle by {\it #1}}}

\def\abstract{\ms\leftskip 3cm\rightskip .5cm\AbstractStyle{\bf \ni Abstract:}\ }
\def\endabstract{\par\leftskip 2cm\rightskip 0cm\NormalStyle\ss}

\SetNumberingModeSN

\def\nab#1{{\buildrel #1\over \na}}
\def\frac[#1/#2]{\hbox{$#1\over#2$}}
\def\Frac[#1/#2]{{#1\over#2}}
\def\({\left(}
\def\){\right)}
\def\[{\left[}
\def\]{\right]}
\def\^#1{{}^{#1}_{\>\cdot}}
\def\_#1{{}_{#1}^{\>\cdot}}
\def\Label=#1{{\buildrel {\hbox{\fiveSerif \ShowLabel{#1}}}\over =}}
\def\<{\kern -1pt}

\def\Dal{\hbox{\tenRelazioni  \char003}}


\def\ExpandAllCNotes{\long\def\CNote##1{%
\BeginSection
	\Note%
 		##1%
	\endNote%
\EndSection%
}}
\ExpandAllCNotes
%
%
%
%


\def\frame#1{\vbox{\hrule\hbox{\vrule\vbox{\kern2pt\hbox{\kern2pt#1\kern2pt}\kern2pt}\vrule}\hrule\kern-4pt}}

\def\Box to #1#2#3{\frame{\vtop{\hbox to #1{\hfill #2 \hfill}\hbox to #1{\hfill #3 \hfill}}}}


\bib{EPS}{J. Ehlers, F.A.E. Pirani, A. Schild, 
{\it The Geometry of Free Fall and Light Propagation},
in General Relativity,  L.OÕRaifeartaigh ed. Clarendon, (Oxford, 1972). 
}

\bib{Perlick}{V. Perlick, 
{\it Characterization of standard clocks by means of light rays and freely falling particles},
General Relativity and Gravitation,  {\bf 19}(11) (1987) 1059-1073}

\bib{BiMetricTheories}{Komar, BiMetricTheories 
}

\bib{nogo}{E. Barausse, T.P. Sotiriou, J.C. Miller,
{\it A no-go theorem for polytropic spheres in Palatini $f(R)$ gravity}, 
Class. Quant. Grav. {\bf 25} (2008) 062001; gr-qc/0703132
}

\bib{Faraoni}{T.P. Sotiriou, V. Faraoni,
{\it  $f (R)$  theories of gravity},
(2008); arXiv: 0805.1726v2
}

\bib{NoGo2}{G.J. Olmo,
{\it  Re-examination of polytropic spheres in Palatini $f(R)$ gravity},
Phys.Rev. D {\bf 78} (2008) 104026; gr-qc/0703132
}

\bib{Capozziello}{S. Capozziello, M. De Laurentis, V. Faraoni
{\it A bird's eye view of $f(R)$-gravity}
(2009); arXiv:0909.4672 
}

\bib{PartI}{M. Di Mauro, L. Fatibene,  M. Ferraris, M. Francaviglia,
{\it Further Extended Theories of Gravitation: Part I};
arXiv:0911.2841
}

\bib{Magnano}{G. Magnano, L.M. Sokolowski, 
{\it On Physical Equivalence between Nonlinear Gravity Theories}
Phys.Rev. D50 (1994) 5039-5059; gr-qc/9312008
}

\bib{Mana}{A. Mana, L. Fatibene, M. Francaviglia
{\it Counter Examples to No-go Theorem  for Polytropic Spheres in Palatini $f(R)$ Gravity}
(in preparation)
}

\bib{Universality}{A. Borowiec, M. Ferraris, M. Francaviglia, I. Volovich,
{\it Universality of Einstein Equations for the Ricci Squared Lagrangians},
Class. Quantum Grav. 15 (1998) 43-55
}

\bib{BiMetricTheories}{Komar, BiMetricTheories 
}

\bib{S1}{T.P. Sotiriou, S. Liberati,
{\it Metric-affine f(R) theories of gravity},
Annals Phys. 322 (2007) 935-966; gr-qc/0604006
}

\bib{S2}{T.P. Sotiriou,
{\it $f(R)$ gravity, torsion and non-metricity},
Class. Quant. Grav. 26 (2009) 152001; gr-qc/0904.2774}

\bib{S3}{T.P. Sotiriou,
{\it Modified Actions for Gravity: Theory and Phenomenology},
Ph.D. Thesis; gr-qc/0710.4438}

\bib{C1}{S. Capozziello, M. Francaviglia,
{\it Extended Theories of Gravity and their Cosmological and Astrophysical Applications},
Journal of General Relativity and Gravitation 40 (2-3), (2008) 357-420.}

\bib{C2}{S. Capozziello, M.F. De Laurentis, M. Francaviglia, S. Mercadante,
{\it From Dark Energy \& Dark Matter to Dark Metric},
Foundations of Physics 39 (2009) 1161-1176
gr-qc/0805.3642v4}

\bib{C3}{S. Capozziello, M. De Laurentis, M. Francaviglia, S. Mercadante,
{\it First Order Extended Gravity and the Dark Side of the Universe: the General Theory}
Proceedings of the Conference ``Univers Invisibile'', Paris June 29 Ð July 3, 2009 
- to appear in 2010}

\bib{C4}{S. Capozziello, M. De Laurentis, M. Francaviglia, S. Mercadante,
{\it First Order Extended Gravity and the Dark Side of the Universe Ð II: Matching Observational Data},
Proceedings of the Conference ``Univers Invisibile'', Paris June 29 Ð July 3, 2009 
Ð to appear in 2010}



\def\ubal{\underline{\al}\kern1pt}
\def\obal{\overline{\al}\kern1pt}

\def\ubR{\underline{R}\kern1pt}
\def\obR{\overline{R}\kern1pt}
\def\ubom{\underline{\om}\kern1pt}
\def\obxi{\overline{\xi}\kern1pt}
\def\ubu{\underline{u}\kern1pt}
\def\ube{\underline{e}\kern1pt}
\def\obe{\overline{e}\kern1pt}

\NormalStyle

\title{Further Extended Theories of Gravitation: Part II\footnote{$^\ast$}{{\AbstractStyle
	This paper is published despite the effects of the Italian law 133/08 ({\tt http://groups.google.it/group/scienceaction}). 
        This law drastically reduces public funds to public Italian universities, which is particularly dangerous for free scientific research, 
        and it will prevent young researchers from getting a position, either temporary or tenured, in Italy.
        The authors are protesting against this law to obtain its cancellation.\goodbreak}}}

\author{L.Fatibene,  M.Ferraris, M.Francaviglia, S. Mercadante}


\abstract
We shall present and analyze two examples of extended theories of gravitation in Palatini formalism with matter that couples to the connection.
This will show that the class of  Further Extended Theories of Gravitation introduced in \ref{PartI} does not trivially reduce to $f(R)$ models.
It will also produce an example of theory that on-shell endowes spacetime with a non-trivial Weyl geometry where the connection is not induced
by the metric structure (though it is compatible with it in the sense of Ehlers-Pirani-Schild; see \ref{EPS}).

\endabstract

\NewSection{Introduction}

In a recent paper \ref{PartI} we introduced the class of Further Extended Theories of Gravitation (FETG) and showed that it encompasses $f(R)$ theories. 
We shall here present two examples of FETG which are not $f(R)$ theories in Palatini framework nor equivalent to them.

These examples could of course be ruled out by some physical principle independent of the EPS axioms, 
or should be analyzed to check whether they could fit observational data.

As in \ref{PartI} $M$ is considered as a connected and paracompact  differential manifold of dimension $4$, which allows global Lorentzian metrics.
Axioms in EPS  (see \ref{EPS}) are assumed to hold and the corresponding Weyl geometry is induced on $M$.

In particular example $2$ will provide an authentic non-trivial example of Weyl geometry endowed naturally by a relativistic field theory;
in fact we shall show that on-shell the connection in that model will be so much as non-metric.
In this model from a kinematical point of view the affine structure of spacetime is determined by the metric structure together with four additional degrees of freedom, hence with more freedom than the conformal freedom obtained in $f(R)$ theories.   
Non-metricity in Palatini framework has been considered (see for example \ref{S2}, \ref{S1}); here, however, the model is considered within FETG framework 
which relies on EPS which enhances physical interpretation of the model by establishing a direct connection with observations.

This also shows that there is no logical obstruction to obtain from a variational principle a non-metric affine structure which is allowed by EPS axioms on a kinematical stance.

\NewSection{Example 1}

Let us first consider on $M$ a metric field $g$, a torsionless connection $\Ga$ and a tensor density $A$ of rank 1 and weight $-1$.
The covariant derivative of $A_\mu$ is then defined as
$$
 \nab{\Ga}_\mu A_\nu= d_\mu A_\nu -\Ga^\la_{\nu\mu} A_\la +\Ga^\la_{\la\mu} A_\nu
\fn$$ 
Accordingly, we have
$$
 \nab{\Ga}_{(\mu} A_{\nu)}=d_{(\mu} A_{\nu)} -\(\Ga^\ep_{\nu\mu}  - \de^\ep_{(\nu}\Ga^\la_{\mu)\la} \)A_\ep=
 d_{(\mu} A_{\nu)} -u^\ep_{\mu\nu}  A_\ep
\fn$$
where we set $u^\ep_{\mu\nu} :=\Ga^\ep_{\mu\nu}  - \de^\ep_{(\mu}\Ga^\la_{\nu)\la}$.

Let us consider the following Lagrangian (density)
$$
L=\frac[1/\ka]\sqrt{g} f(R) + g g^{\mu\nu} \nab{\Ga}_\mu A_\nu
\fn$$
where $g=|\det(g_{\mu\nu})|$, $R=g^{\mu\nu} R_{\mu\nu}(\Ga)$ is the scalar curvature of $(g, \Ga)$, $\ka=16\pi G$ is a constant and 
$f$ a generic (analytic) function; see \ref{Universality}.

By variation of this Lagrangian and usual covariant integration by parts one obtains
$$
\eqalign{
\de L&=
 \frac[\sqrt{g}/\ka]\( f'(R) R_{(\al\be)} -\frac[1/2] f(R) g_{\al\be} -\ka T_{\al\be }\)\de g^{\al\be}+\cr
& - g g^{\al\be}  A_\la  \de u^\la_{\al\be}+\frac[\sqrt{g}/\ka] g^{\al\be} f'(R) \nab{\Ga}_{\la} \de u^\la_{\al\be} 
+ g  g^{\mu\nu} \nab{\Ga}_\mu \de A_\nu =\cr
=&
\frac[\sqrt{g}/\ka] \( f'(R) R_{(\al\be)} -\frac[1/2] f(R) g_{\al\be} -\ka T_{\al\be }\)\de g^{\al\be} 
-\frac[1/\ka]\(\nab{\Ga}_{\la}\(\sqrt{g} g^{\al\be} f'(R)\)+\ka g g^{\al\be}  A_\la  \)  \de u^\la_{\al\be}+\cr
& 
 - \nab{\Ga}_\mu\(g  g^{\mu\nu}\)  \de A_\nu 
+ \nab{\Ga}_\la \( \frac[\sqrt{g}/\ka] g^{\al\be} f'(R) \de u^\la_{\al\be}+g  g^{\la\nu}  \de A_\nu\) \cr
}
\fn$$
where we used the well-known identity $\de R_{(\al\be)}=\nab{\Ga}_\la \de u^\la_{\al\be}$
and 
we set for the energy-momentum tensor $ T_{\al\be}:= \sqrt{g}\( g_{\al\be} g^{\mu\nu} \nab{\Ga}_\mu A_\nu - \nab{\Ga}_{(\al} A_{\be)}\)$.

Field equations are
$$
\cases{
&f' R_{(\al\be)} -\frac[1/2] f g_{\al\be} =\ka T_{\al\be }\cr
&\nab{\Ga}_{\la}\(\sqrt{g} g^{\al\be} f'\)=\al_\la \sqrt{g} g^{\al\be}f'  \cr
& \nab{\Ga}_\mu\(g  g^{\mu\nu}\)=0
}
\fn$$
where we set $\al_\la:=-\ka \frac[ \sqrt{g}/f'] A_\la$.
Notice that the third equation (that is the matter field equation) is not enough to fix the connection due to the contraction.
Notice also that these are more general than field equations of standard $f(R)$ theories due to the rhs of the second equation
(that is originated by the coupling between the matter field $A$ and the connection $\Ga$).
Nevertheless one can analyze these field equations along the same lines used in $f(R)$ theories.
Let us thence define a metric $h_{\mu\nu}= f'  g_{\mu\nu}$ and rewrite the second equation as
$$
\nab{\Ga}_{\la}\(\sqrt{h} h^{\al\be} \)=\al_\la \sqrt{h} h^{\al\be}
\fn$$
According to the analysis of EPS-compatibility done in \ref{PartI} this fixes the connection as
$$
\Ga^\al_{\be\mu}:=  \{ h \}^\al_{\be\mu}  - \frac[ \ka /2f'] \( h^{\al\ep}h_{\be\mu} - 2\de^\al_{(\be} \de^\ep_{\mu)} \) a_\ep 
\fn$$ 
where for notational convenience we introduced the $1$-form $a_\ep:=\sqrt{g}A_\ep$.
For later convenience let us notice that we have
$$
K^\al_{\be\mu}\equiv \Ga^\al_{\be\mu}- \{ h \}^\al_{\be\mu}= -\frac[ \ka /2f'] \( h^{\al\ep}h_{\be\mu} - 2\de^\al_{(\be} \de^\ep_{\mu)} \) a_\ep 
\fn$$

Now we can define the tensor $H^\al_{\be\mu}:=\Ga^\al_{\be\mu}-\{g\}^\al_{\be\mu}$ and obtain
$$
H^\al_{\be\mu}=
K^\al_{\be\mu}- \frac[1/2 ] \( g^{\al\la}g_{\be\mu}-2 \de^\la_{(\be}\de^\al_{\mu)}  \)\del_{\la} \ln f'=
- \frac[1/2f' ] \( g^{\al\ep}g_{\be\mu}-2 \de^\ep_{(\be}\de^\al_{\mu)}  \)\( \ka a_\ep +\del_{\ep} f'\)
\fn$$ 
By substituting into the third field equation
we obtain
$$
\eqalign{
\nab{g}_\mu&\(g  g^{\mu\nu}\) + g\( H^\mu_{\la\mu} g^{\la\nu}+ H^\nu_{\la\mu}  g^{\mu\la}- 2H^\la_{\la\mu} g^{\mu\nu}\)=0\cr
&\then  H^\nu_{\la\mu}  h^{\mu\la}-H^\la_{\la\mu} h^{\mu\nu}=0\cr
&\then 
- \frac[1/2f' ] \( \(h^{\nu\ep}h_{\la\mu}-2 \de^\ep_{(\la}\de^\nu_{\mu)}\)h^{\mu\ep}- \(h^{\la\ep}h_{\la\mu}-2 \de^\ep_{(\la}\de^\la_{\mu)}\) h^{\mu\nu} \)\( \ka a_\ep +\del_{\ep} f'\)=0\cr
&\then
- \frac[3/f' ] h^{\nu\ep} \( \ka a_\ep +\del_{\ep} f'\)=0
\quad\then
 a_\ep =-\frac[1/\ka]\del_{\ep}  f' 
\cr
}
\fn$$
where $\nab{g}_\mu$ is now the covariant derivative with respect to the metric $g$. 
Hence the matter field $A_\ep=\sqrt{g}a_\ep= -\frac[\sqrt{g}/\ka]\del_{\ep}  f' $ has no dynamics and it is completely determined in terms of the other fields.

We can also express the connection as a function of $g$ alone (or, equivalently, of $h$ alone)
$$
\Ga^\al_{\be\mu}:=  \{ h \}^\al_{\be\mu}  + \frac[ 1 /2] \( h^{\al\ep}h_{\be\mu} - 2\de^\al_{(\be} \de^\ep_{\mu)} \) \del_{\ep}  \ln f' 
\equiv  \{ g \}^\al_{\be\mu} 
\fn$$
This behaviour, which has been introduced by the matter coupling, is quite peculiar; the model resembles in the action an $f(R)$ theory but in solution space
the connection is directly  determined by the original metric rather than by the conformal metric $h$ as in $f(R)$ theories. Still the metric $g$ obeys modified
Einstein equations. 
In fact, we have the first field equation which is now depending on $g$ alone, since the matter and the connection have been determined as functions of $g$.

The {\it master equation} is obtained as usual by tracing (using $g^{\al\be}$)
$$
f' R -2 f  =\ka T
\qquad\then
f=\frac[1/2]\(f'R-\ka T\)
\fn$$
where we set $T:= T_{\al\be} g^{\al\be}$.
Notice that in this case we obtain explicitly
$$
\eqalign{
&T_{\al\be}=\frac[1/\ka]\( \na_{(\al} \na_{\be)}f'-g_{\al\be} \Dal f' \)\cr
&T= 4\na_{\al} a_{\be}g^{\al\be}-g^{\mu\nu} \na_\mu a_\nu = -\frac[3/\ka] \Dal  f'  
}\fn$$
The master equation is then  $f' R -2 f  = -3\Dal  f' $.
Then substituting back into the first field equation we obtain
$$
\eqalign{
&f' \(R_{\al\be} - \frac[1/4] Rg_{\al\be}\)  -\frac[3/4] \Dal f' g_{\al\be} =  \na_{(\al} \na_{\be)}f'- \Dal f' g_{\al\be}\cr
&\quad\then
R_{\al\be} - \frac[1/2] Rg_{\al\be}  = \frac[1/f']  \(\na_{(\al} \na_{\be)}f'-\frac[1/4] \(\Dal f' +  f' R\) g_{\al\be}  \)
}
\fn$$
where now the curvature and covariant derivatives refer to $g$. 
These are exactly the field equations obtained in the corresponding purely-metric $f(R)$ theory.

Hence we have that, regardless of the function $f$, when there is no matter field other than the field $A$  all these models behave 
exactly as metric $f(R)$ theories.
Unlike in $f(R)$ theories, however, there is no conformal metric around; everything refers to the original metric $g$.  

Obviously in this theory one can use the purely metric model for polytropic star as a solution to find a possible way around the
no-go theorems formulated for Palatini extended theories; see \ref{nogo}. Another possible way around will be presented in \ref{Mana}.

\NewSection{Example 2}

The analysis of Example 1 is based on the assumption that the matter field $A_\ep$ is fundamental (or equivalently that $\de A_\ep$ are independent of other field variations); on the other hand the EPS-compatibility is based on the geometric character of the matter field $A_\ep$.
Let us now consider what happens when the tensor density $A_\ep$ is obtained as an object  derived from $g$ and other, more fundamental, matter fields.
Let us for example consider a (real) scalar field $\phi$ and set $A_\ep= \frac[1/\sqrt{g}] \na_\ep\phi$. 
(Notice that the covariant derivative of $\phi$ is in fact independent of any connection since for scalars $\na_\mu\equiv \del_\mu$.)

Accordingly, let us consider a second model with the Lagrangian
$$
L=\frac[1/\ka]\sqrt{g} f(R) + g g^{\mu\nu} \nab{\Ga}_\mu \(\frac[1/\sqrt{g}] \na_\nu\phi\)
\fn$$
By variation of this Lagrangian and usual covariant integration by part one obtains
$$
\eqalign{
\de L&=
\frac[\sqrt{g}/\ka] \( f'(R) R_{(\al\be)} -\frac[1/2] f(R) g_{\al\be} -\ka T_{\al\be }\)\de g^{\al\be} 
+\cr
& 
-\frac[1/\ka]\(\nab{\Ga}_{\la}\(\sqrt{g} g^{\al\be} f'(R)\)+\ka g g^{\al\be}  A_\la  \)  \de u^\la_{\al\be}
 +\na_\nu \( \frac[1/\sqrt{g}] \nab{\Ga}_\mu\(g  g^{\mu\nu}\)\)  \de\phi +\cr
& + \nab{\Ga}_\la \( \frac[\sqrt{g}/\ka] g^{\al\be} f'(R) \de u^\la_{\al\be}
 +\frac[ \sqrt{g}/2]  g^{\la\nu} g_{\al\be} \na_\nu \phi   \de g^{\al\be} +\sqrt{g}  g^{\la\nu} \na_\nu \de\phi 
 - \frac[1/\sqrt{g}] \nab{\Ga}_\mu\(g  g^{\mu\la}\) \de\phi\) \cr
}
\fn$$
where we set $ T_{\al\be}:= \sqrt{g}\( g_{\al\be} g^{\mu\nu} \nab{\Ga}_\mu A_\nu - \nab{\Ga}_{(\al} A_{\be)}\) -\frac[1/2 g] \nab{\Ga}_\mu\(g  g^{\mu\nu}\) \na_\nu  \phi g_{\al\be}$.
Here we denote by $\nab{\Ga}_\mu$ the covariant derivative wrt the connection $\Ga$, while $\na_\mu$ is used for the special cases in which the covariant derivative turns out to be independent of any connection and reduces to a partial derivative (as it happens for scalars, vector densities of weight $1$, and so on). 

Field equations are
$$
\cases{
&f' R_{(\al\be)} -\frac[1/2] f g_{\al\be} =\ka T_{\al\be }\cr
&\nab{\Ga}_{\la}\(\sqrt{g} g^{\al\be} f'\)=\al_\la \sqrt{g} g^{\al\be}f'  \cr
& \na_\nu \( \frac[1/\sqrt{g}] \nab{\Ga}_\mu\(g  g^{\mu\nu}\)\)  =0
}
\fn$$
where we set again $\al_\la:=-\ka \frac[ \sqrt{g}/f'] A_\la$.
The second equation fixes again the connection
$$
\Ga^\al_{\be\mu}:=  \{ h \}^\al_{\be\mu}  - \frac[ \ka /2f'] \( h^{\al\ep}h_{\be\mu} - 2\de^\al_{(\be} \de^\ep_{\mu)} \) \na_\ep\phi 
\fn$$ 
where as usual we set $h_{\mu\nu}= f' g_{\mu\nu}$.

However, the third equation does not force the covector $\al_\ep=-\frac[ \ka /f'] \na_\ep\phi $ to be
 a closed form; thus the connection is not metric.

\Note
To see this, notice that the third equation is in the form $d\ast \be =0$ for a covector $\be= \be_\nu dx^\nu$. 
Here $\ast$ denotes the Hodge duality on forms.
In fact, the third equation can be recasted as 
$$
\na_\nu \( -\Frac[3\sqrt{g}/f' ] g^{\nu\ep} \na_\ep \(\ka\phi + f'\)\)=0
\qquad \then
\ast \be=  -\Frac[3\sqrt{g}/f' ] g^{\nu\ep} \na_\ep \(\ka\phi + f'\) ds_\nu
\fn$$

The general solution of this equation is
$$
\ast \be= d \te  + \om
\fn$$
for a closed $(m-1)$-form $\om=\om^\mu ds_\mu$ and for some $(m-2)$-form $\te=\frac[1/2]\sqrt{g}\>\te^{\mu\nu} ds_{\mu\nu}$.
The closed form $\om$ is defined modulo exact forms and they are classified in terms of spacetime cohomology.

Accordingly, the third equation implies
$$
\ka\na_\ep\phi = \frac[1/3] g_{\ep\nu} \frac[f'/ \sqrt{g}] \[\na_\la\(\sqrt{g} \te^{\la\nu}\)	+\om^\nu\] -\na_\ep f'  
\fn$$
Consequently, 
$$
\eqalign{
\Ga^\al_{\be\mu}:=&  \{ h \}^\al_{\be\mu}  
-\frac[1/6] \frac[ 1 /\sqrt{g}] \( g^{\al\ep}g_{\be\mu} - 2\de^\al_{(\be} \de^\ep_{\mu)} \)  g_{\ep\nu}\[\na_\la\(\sqrt{g} \te^{\la\nu}\) +\om^\nu \]+\cr
&+ \frac[ 1 /2] \( g^{\al\ep}g_{\be\mu} - 2\de^\al_{(\be} \de^\ep_{\mu)} \)\na_\ep \ln f'  
}
\fn$$ 
which corresponds to
$$
\al_\ep=\na_\ep \ln f'  -  \frac[1/3] g_{\ep\nu}\frac[ 1 /\sqrt{g}] \[\na_\la\(\sqrt{g} \te^{\la\nu}\) +\om^\nu	\]
\fn$$

The connection $\Ga$ is metric iff the covector $\al=\al_\ep dx^\ep$ is closed. However, there is nothing here forcing this form to be closedt
(while of course it can be closed for specific choices of the arbitrary $\te^{\mu\nu}$, e.g.~$\te=0$).
For example, if $g$ is Minkowski metric, $\om=0$ and $\te= \sqrt{g} (x^1)^2 ds_{12}$ one can prove that $d\al\not=0$ holds.

\endNote

The field $A$ can thence be written as
$$
A_\ep= \Frac[1/\sqrt{g}]  \na_\ep \phi=   \Frac[f'/3\ka g] g_{\ep\nu}  \na_\la\(\sqrt{g} \te^{\la\nu}\) -\Frac[1/\ka \sqrt{g}]\na_\ep f'  
\fn$$
Let us stress that now the matter field $\phi$ is not completely determined by the other fields (there is in fact a freedom 
in the choice of the form $\te^{\mu\nu}$).


The master equation induced by the first field equation is in this case
$$
f'  R -2 f = \ka T_{\al\be} g^{\al\be}=: \ka T   
\qquad \then
f= \frac[1/2]\(f'  R - \ka T\)
\fn$$
which can be used back into the first field equation to obtain (when $f'\not=0$)
$$
R_{(\al\be)} - \frac[1/4] R g_{\al\be}= \frac[\ka/f'] \( T_{\al\be} + \frac[1/2] T g_{\al\be}\)
\fn$$

Similar examples are obtained any time that one can define a tensor density $A_\ep$ of weight $-1$ from any choice of fundamental fields.

\NewSection{Conclusions and Perspectives}

We do not pretend here to propose any realistic physical model. 
In order to do that one should study specific models; for example in their cosmological mini--superspace or other astrophysical situations
and try fitting observational data; see \ref{C1}, \ref{C3}, \ref{C4}, \ref{S3}.

We are here just considering the possibility to use EPS compatibility in order to constrain extended theories of gravitation.
Since EPS criteria allow for non-metric connections it is interesting to notice that in fact a specific model (Example $2$) can be presented in which
non-metric connections appear naturally.

Of course,  these examples are defined {\it ad hoc} and may have no physical meaning whatsoever; however, this is hard to be seen as a critic. In fact once one accepts to introduce exotic 
dynamics (if not even considering Hilbert-Einstein gravitation as a {\it special} model) then it is difficult to
set a point not to be crossed and any exotic model should be discussed in view of its own prediction.
From this point of view it is quite interesting to notice that EPS criteria are a natural crosspoint to unphysical models.
EPS axioms are quite concrete and physically well-based. This does not imply  of course that they are a complete set of hypotheses.
There could be further reasons to exclude the models we presented here, possibly by adding new criteria to what should be meant 
by ``physical connections''.
However, in this case such principles should be explicitly formulated and discussed.

Moreover, EPS setting provides a natural framework for relativistic theories of gravitation in Palatini formalism.
Let us stress that further investigations are needed in order to provide a truly relativistic operational definition of measurements
in this generalized setting; see \ref{C2}, \ref{Perlick} 

\Acknowledgements

This work is partially supported by MIUR: PRIN 2005 on {\it Leggi di conservazione e termodinamica in meccanica dei continui e teorie di campo}.  
We also acknowledge the contribution of INFN (Iniziativa Specifica NA12) and the local research funds of Dipartimento di Matematica of Torino University.

\ShowBiblio

\end